\documentclass[12pt,a4paper]{article}
\usepackage{amsmath,amssymb}
\usepackage{epsfig,graphicx}
\usepackage{subfigure}
\usepackage{graphicx}
\usepackage{bm}
\usepackage{color}
\usepackage{axodraw}
\usepackage[sort&compress,square]{natbib}
\usepackage[scale={0.8,0.83},vmarginratio=8:10,hmarginratio=1:1,headheight=10pt,footskip=20pt]{geometry}

\makeatletter
\@addtoreset{equation}{section}
\makeatother

\newcounter{multieqs}

\newcommand{\imi}{\mathrm{i}}

\begin{document}

\title{\vspace*{-2cm}
\hfill{\small DCPT/08/110}\\[-0.3cm]
\hfill{\small DESY 08-105}\\[-0.3cm]
\hfill{\small IPPP/08/55}\\[-0.3cm]
\hfill{\small OUTP-0810P}\\
\vspace{20pt} \Large{\textbf{{Probing Hidden Sector Photons through the \mbox{Higgs Window}}}} }

\author{
Markus Ahlers$^{1,}$\footnote{{\bf
e-mail}: m.ahlers1@physics.ox.ac.uk}\,\,, Joerg Jaeckel$^{2,}$\footnote{{\bf
e-mail}: joerg.jaeckel@durham.ac.uk}\,\,, Javier Redondo$^{3,}$\footnote{{\bf
e-mail}: javier.redondo@desy.de}\,\,, and Andreas Ringwald$^{3,}$\footnote{{\bf
e-mail}: andreas.ringwald@desy.de}
\\[2ex]
\small{\em $^1$Rudolf Peierls Centre for Theoretical Physics, University of Oxford, Oxford OX1 3NP, UK}\\[-0.1cm]
\small{\em $^2$Institute for Particle Physics and Phenomenology, Durham University, Durham DH1 3LE, UK}\\[-0.1cm]
\small{\em $^3$Deutsches Elektronen-Synchrotron, Notkestra\ss e 85, 22607 Hamburg, Germany}
}

\date{}

\maketitle
\vspace{-0.5cm}
\begin{abstract}
\noindent
We investigate the possibility that a (light) hidden sector extra photon receives its mass via spontaneous symmetry breaking of a hidden sector Higgs boson, the so-called \emph{hidden-Higgs}. The hidden-photon can mix with the ordinary photon via a gauge kinetic mixing term. The hidden-Higgs can couple to the Standard Model Higgs via a renormalizable quartic term -- sometimes called the \emph{Higgs Portal}. We discuss the implications of this light hidden-Higgs in the context of laser polarization and light-shining-through-the-wall experiments as well as cosmological, astrophysical, and non-Newtonian force measurements. For hidden-photons receiving their mass from a hidden-Higgs we find in the small mass regime significantly stronger bounds than the bounds on massive hidden sector photons alone.
\end{abstract}

\section{Introduction}

Many extensions of the Standard Model contain so-called hidden-sectors which interact only very weakly with the {known particles from the visible sector}. Due to their feeble interactions particles in these hidden sectors can be easily missed in conventional collider experiments. Therefore, the bounds on their masses are often very weak and even masses in the sub-eV regime are possible. Such small masses, however, open the possibility that these particles may be detectable in low energy high precision experiments. Moreover, they could leave observable footprints in astrophysics and cosmology. This could therefore open a new window into particle physics which could give us crucial complementary information about the underlying laws of nature.

\begin{figure}[tb]
\begin{center}
\includegraphics[width=0.7\linewidth]{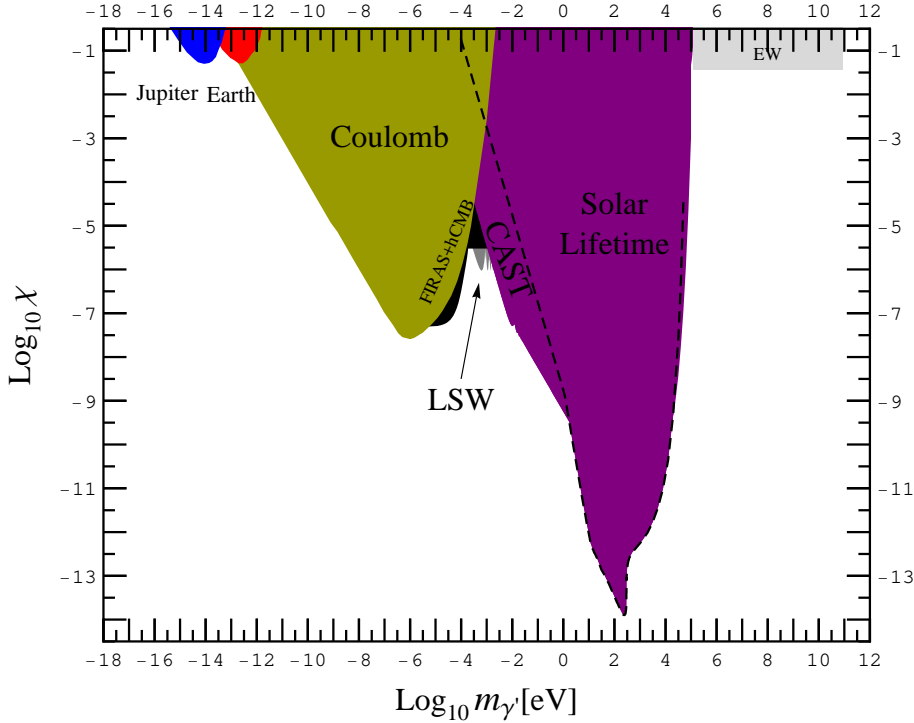}
\end{center}
\vspace{-2ex} \caption[...]{\small Current bounds on hidden-sector
photons from analyzing the magnetic fields of Jupiter and Earth~\cite{Goldhaber:1971mr}, Coulomb law tests
\cite{Williams:1971ms,Bartlett:1988yy} (gold), electroweak precision
data \cite{Feldman:2007wj} (lightgray), searches of solar
hidden-photons with the CAST experiment (purple)
\cite{Popov:1991,Popov:1999,Andriamonje:2007ew,Redondo:2008aa} and
light-shining-through-walls (LSW) experiments
\cite{Cameron:1993mr,Robilliard:2007bq,Chou:2007zz,Ahlers:2007rd,Ahlers:2007qf}
{(grey)} as well as CMB measurements of the effective number of
neutrinos and the blackbody nature of the spectrum (black)
\cite{Mangano:2006ur,Ichikawa:2006vm,Komatsu:2008hk,Jaeckel:2008fi}.
Improvements of the solar bounds can be achieved using the
SuperKamiokande detector or upgrading the CAST experiment
\cite{Gninenko:2008pz}. The region $m_{\gamma^{\prime}}\lesssim {\rm
few}\,\,{\rm meV}$ could be tested by an experiment using microwave
cavities \cite{Jaeckel:2007ch,Penny} or experiments searching for
magnetic fields leaking through a superconducting
shielding~\cite{Jaeckel:2008sz}. \label{bounds1}}
\end{figure}

One interesting class of hidden sector particles is additional U(1) gauge bosons, {\it i.e.}~hidden-photons. For example many models arising from string compactifications contain extra U(1) gauge particles under which the Standard Model particles are uncharged. Accordingly the only renormalizable interaction of the hidden-photon with the Standard Model is via mixing of the hidden-photon with the ordinary electromagnetic photon~\cite{Okun:1982xi,Holdom:1985ag,Foot:1991kb}. Current constraints on this mixing are shown in Fig.~\ref{bounds1}.

As can be seen from Fig.~\ref{bounds1} the bounds depend crucially on the mass of the extra photon. In particular for very small masses the bounds become very weak. In this note we want to investigate if knowledge of the mechanism which generates a mass for the hidden-photon can improve the bounds. In principle a mass for the hidden-photon can be generated either via a Higgs mechanism or via a St\"uckelberg mechanism \cite{Stueckelberg:1938}.
Here, we focus mainly on the case of the Higgs mechanism.

The crucial difference between the Higgs and the St\"uckelberg mechanism is that the gauge boson acquires a mass from the expectation value of a physical boson. As we will see in the following this additional boson will open new avenues of detection. Moreover, the additional physical boson also allows for a new possible renormalizable coupling to the Standard Model. The hidden sector Higgs can mix with the Standard Model Higgs via a quartic term~\cite{Foot:1991bp}, sometimes called the Higgs-Portal~\cite{portalref}.

The paper is organized as follows. In the next Sect.~\ref{gauge} we
will present the essentials of the hidden-photon hidden-Higgs system
including a gauge kinetic mixing term with the photon. This will
already lead us to our first main conclusion. In processes where the
momentum transfer is greater than the mass of the hidden-photon the
hidden-Higgs behaves essentially like a minicharged particle and
corresponding (strong) astrophysical and cosmological bounds apply.
Then in Sect.~\ref{magnetic} we discuss the effects of a strong
magnetic field as relevant for laser polarization and
light-shining-through-walls experiments. Again we find that the
bounds improve significantly. Moreover we suggest possible ways to
experimentally distinguish between the Higgs and St\"uckelberg
mechanisms. In Sect.~\ref{Portal} we then include effects of
electroweak symmetry breaking and a possible mixing of the
hidden-Higgs with the Standard Model Higgs via a Higgs-Portal term.
Bounds from fifth-force experiments provide interesting constraints
on the Higgs-Portal term which are independent of the size of the
kinetic mixing. Finally, in Sect.~\ref{conclusions} we summarize and
conclude.

\section{Gauge Kinetic Mixing and the Hidden-Higgs}\label{gauge}

The most general kinetic term of the $\mathrm{U}(1)_\text{EM}\times \mathrm{U}(1)_X$ gauge fields includes a kinetic mixing term of the corresponding field tensors
\begin{equation}
\mathcal{L}^{\mathrm{U}(1)}_\text{kin} = -\frac{1}{4}F^{\mu\nu}F_{\mu\nu}-\frac{1}{4}X^{\mu\nu}X_{\mu\nu} -\frac{\chi}{2}F^{\mu\nu}X_{\mu\nu}\,,
\end{equation}
where $F^{\mu\nu}$ is the ordinary electromagnetic field strength and $X^{\mu\nu}$ is the field strength corresponding to the hidden sector gauge field $X^{\mu}$~\footnote{In previous publications we have often used $B^{\mu}$ to denote the hidden sector gauge field. However, since we will later consider mixing with the hypercharge which is conventionally denoted by $B^\mu$ we will switch to $X^{\mu}$.}. The first two terms are the ordinary kinetic terms for the fields $A^{\mu}$ and $X^{\mu}$, respectively. The last term is the kinetic mixing~\cite{Holdom:1985ag}. This term is a renormalizable dimension four operator and consequently $\chi$ does not suffer from mass suppressions. It is therefore a sensitive probe for physics at very high energies. Kinetic mixing arises in field theoretic \cite{Holdom:1985ag} as well as in string theoretic setups \cite{stringref}. Typical predicted values for $\chi$ in realistic string compactifications range between $10^{-16}$ and $10^{-2}$.

This kinetic mixing term can be diagonalized by a shift of the $X_\mu$ term and a redefinition of the electromagnetic coupling as
\begin{equation}
\label{shift}
X_\mu \to X_\mu - \chi A_\mu\quad{\rm and}\quad e^2\to
\frac{e^2}{1-\chi^2}\,.
\end{equation}
After the shift the covariant derivative takes the form
\begin{equation}
D_\mu = \partial_\mu-\imi(Q_\text{EM}e-\chi Q_X g_X)A_\mu-\imi Q_X g_X X_\mu\,.
\end{equation}
A particle charged only under the ordinary electromagnetic U(1), $Q_{X}=0$, remains unchanged. However, a hidden sector particle with $Q_{X}\neq 0$ and $Q_\text{EM}=0$, receives a small electric charge,
\begin{equation}
\label{charge}
{Q_{\text{EM}, \text{mixing}}=-\chi\left(\frac{g_{X}}{e}\right)Q_{X}}\,,
\end{equation}
under the ordinary electromagnetic U(1).

In Eq.~\eqref{charge} not only the kinetic mixing parameter but also the hidden sector gauge coupling appears. From a pure field theoretical point of view it would be natural to assume that $g_X\sim e$. However, in some string scenarios with large volumes of the extra dimensions the couplings could be reduced by a large factor of order $10^{-4}$ or smaller~\cite{Burgess:2008ri}.

\begin{figure}[t]
\begin{center}
\scalebox{0.75}[0.75]{
\begin{picture}(250,130)(110,0)
\Photon(0,20)(00,100){5}{7.5}
\Text(-35,60)[l]{\scalebox{1.333}[1.333]{$A^{\mu}$}}
\Text(0,120)[c]{\scalebox{1.333}[1.333]{$j^{\mu}_{h}$}}
\LongArrow(-18,108)(18,108)
\Text(150,120)[c]{\scalebox{1.333}[1.333]{$j^{\mu}_{h}$}}
\LongArrow(132,108)(168,108)
\Vertex(0,100){3}
\Text(30,90)[c]{\scalebox{1.333}[1.333]{$-\chi g_{X}$}}
\Line(-45,100)(0,100)
\Line(0,100)(45,100)
\Text(75,75)[c]{\scalebox{2.0}[2.0]{$+$}}
\Photon(150,20)(150,100){5}{7.5}
\Text(165,60)[l]{\scalebox{1.333}[1.333]{$\chi m^2_{\gamma^{\prime}}$}}
\Text(165,90)[l]{\scalebox{1.333}[1.333]{$g_X$}}
\Text(115,40)[l]{\scalebox{1.333}[1.333]{$A^{\mu}$}}
\Text(115,80)[l]{\scalebox{1.333}[1.333]{$X^{\mu}$}}
\Vertex(150,100){3}
\Line(105,100)(150,100)
\Line(150,100)(195,100)
\Text(220,65)[l]{\scalebox{1.333}[1.333]{$\displaystyle\sim-\chi g_{X}\frac{q^2}{q^2-m^{2}_{\gamma^{\prime}}}=\bigg\{\begin{array}{ccc}
                                                                                                        0 & {\rm for} & q^2=0 \\
                                                                                                        -\chi g_{X} & {\rm for} & |q^2|\gg m^{2}_{\gamma^{\prime}}
                                                                                                      \end{array}$}}
\SetWidth{1.5}
\Line(140,70)(160,50)
\Line(160,70)(140,50)
\end{picture}
}
\end{center}
\vspace{-1.0cm} \caption{\small Contributions to the coupling of the photon to the current generated by a hidden-sector particle $h$ in a situation where the hidden-photon is massive. The first is the direct contribution via the charge {$Q_{\rm EM, mixing}\, e$} that arises from the shift \eqref{shift} of the hidden-photon field. The second is due to the $A^{\mu}-X^{\mu}$ oscillations caused by the non-diagonal mass term \eqref{massterm}. Note that the second diagram is only present if the hidden-photon has non-vanishing mass $m^{2}_{\gamma^{\prime}}\neq 0$.}
\label{interaction}
\end{figure}

Let us now turn to the hidden-Higgs. The hidden sector gauge group can be Higgsed by a particle $\theta$ with charges $(0,q_X\neq 0)$ under the visible and hidden sector U(1), respectively. For a suitable potential,
\begin{equation}
V_\text{Higgs}=-\mu^{2}_{\theta}|\theta|^2+\lambda_{\theta}|\theta|^4,
\end{equation}
the hidden-Higgs aquires a vacuum expectation value
\begin{equation}
\langle\theta\rangle=\frac{v_{\theta}}{\sqrt{2}}=\frac{1}{\sqrt{2}}\sqrt{\frac{\mu^2_{\theta}}{\lambda_{\theta}}}\,.
\end{equation}
For the hidden sector gauge field this then results in a mass term,
\begin{equation}
\label{massterm}
{\mathcal {L}}_\text{mass}={\frac{v^{2}_{\theta}}{2}q_X^2g^{2}_{X} X^{\mu}X_{\mu}
{\xrightarrow[{\rm \,\,Eq.\,\,\eqref{shift}}]{{\rm shift}}}\frac{v^{2}_{\theta}}{2}q_X^2g^{2}_{X} (X^{\mu}X_{\mu}-2\chi X^{\mu}A_{\mu}+\chi^2 A^{\mu}A_{\mu})}\,.
\end{equation}
This (non-diagonal) mass-term now leads to the familiar photon--hidden-photon oscillations~\cite{Okun:1982xi}.

The bounds on hidden-photons derived in previous publications~\cite{Redondo:2008aa,Ahlers:2007qf,Ahlers:2007rd} relied solely on the mass term \eqref{massterm}. However, if this term arises from a Higgs mechanism we still have to consider the \emph{real} dynamical Higgs field $\theta$, defined via the replacement\footnote{For simplicity, we work in the unitary gauge in the following.}
\begin{equation}
\theta \to \frac{1}{\sqrt{2}}\left(v_\theta+\theta\right)\,.
\end{equation}
In the symmetric phase where $v_{\theta}=0$ the hidden-Higgs which {has initially only a charge $q_X$} under $\mathrm{U}(1)_X$ receives a fractional electric charge
\begin{equation}
\label{higgscharge}
q_{\theta}=-\chi\left(\frac{g_X}{e}\right){q_X}\,.
\end{equation}
In the spontaneously broken phase things are slightly more tricky. Due to the non-diagonal mass term in Eq.~\eqref{massterm} additional diagrams arise. This is shown in Fig.~\ref{interaction}. For small momenta of the initial photon these diagrams exactly cancel the minicharge. This is to be expected because the hidden-Higgs (or any other hidden sector particle) only couples to the now massive gauge boson $X^{\mu}$ and the interaction mediated by this particle is switched off by its mass\footnote{Another argument is the following. Higgs particles do not couple to a state which remains massless.}. At momentum transfer higher than the mass, $|q^2|\gg m^{2}_{\gamma^{\prime}}$, the interaction mediated by the massive particle $X^{\mu}$ cannot be neglected and the full minicharge is effective (cf.~Fig.~\ref{interaction}).

This gives our first result. In processes in which the momentum transfer is high compared to the mass of the hidden sector photon we have to take into account that the Higgs particle acts as an extra minicharged particle (MCP) (this has already been noticed in~\cite{Melchiorri:2007sq}). This is in particular the case in many astrophysical and cosmological environments. Accordingly, we can simply translate the bounds on minicharged particles into bounds on the kinetic mixing of photons which acquire their mass via a Higgs mechanism, if $m_{\gamma^{\prime}}\lesssim T_\text{bound}$, the temperature of the astrophysical object from which the bound originates. For the sun and red giants this is of the order of ${\rm few}\times{\rm keV}$ and ${\rm few}\times 10\,{\rm keV}$, respectively. This leads to the very strong bounds
\begin{equation}
\chi \lesssim 10^{-14}\left(\frac{e}{g_{X}}\right)\quad{\rm for}\quad m_{\gamma^{\prime}}\lesssim {\rm few}\,{\rm keV}.
\end{equation}
In Fig.~\ref{newbounds} we have made a compilation of the different bounds that arise from the
minicharged Higgs (cf. \cite{Davidson:1993sj,Davidson:2000hf,Badertscher:2006fm}).

\section{External Magnetic Field}\label{magnetic}

A crucial question for our investigation is whether the hidden-Higgs vacuum expectation value actually persists inside a strong magnetic field. The field equation for $\theta$ reads,
\begin{equation}
\label{landau}
\left[\partial^{\mu}-\imi {q_{X}} g_{X} ({X}^{\mu}-\chi A^{\mu})\right]^2\theta+V'(\theta)=0.
\end{equation}
For a static homogenous magnetic field (pointing in the x-direction) inside a coil we can write the external vector potential as
\begin{equation}
\label{pot}
\mathbf{A}_B = \frac{1}{2}\mathbf{B}\times\mathbf{r} = \frac{1}{2}\begin{pmatrix}0\\-zB\\yB\end{pmatrix}\,.
\end{equation}
In the vicinity of the source and in absence of hidden sector sources we have {${X}^{\mu}=0$}.

To check whether we expect a non-vanishing vacuum expectation value for $\theta$ we can neglect the stabilizing $\lambda_{\theta} |\theta|^4$ terms in the potential and search for tachyonic modes. The simplified equation of motion reads,
\begin{equation}
\left[\partial^{\mu}+\imi  q_{\theta} e (A^{\mu}_{\mathbf{B}})\right]^2\theta-\mu^2_{\theta}\theta=0\,,
\end{equation}
where we have used the expression \eqref{higgscharge} for the electric charge of the hidden-Higgs. This is simply the quantum mechanical problem of a particle of charge $q_{\theta} e$ moving in a constant magnetic field $B$. The solution are the famous Landau levels,
\begin{equation}
{\omega^{2}_{n}=}-\mu^2_{\theta}+p^{2}_{x}+2|e q_{\theta} B| \left(n+\frac{1}{2}\right)\quad n=0,1,2,\ldots\,,
\end{equation}
where $n$ is the number of the Landau level and $p_{x}$ is the momentum in the x-direction. The lowest $n=0,p_{x}=0$ mode has
\begin{equation}
\omega^2_{0}=-\mu^{2}_{\theta}+\left|q_{\theta} e B\right|\,.
\end{equation}
This mode is manifestly real and therefore non-tachyonic if
\begin{equation}
\left|q_{\theta} e B\right|\geq \mu^{2}_{\theta}\,.
\end{equation}
In other words for strong enough magnetic fields the hidden sector U(1) symmetry is unbroken\footnote{This effect is similar to the breakdown of superconductivity in strong magnetic fields.}. Hence, if $|q_{\theta} e B|>\mu_\theta^2$ the hidden $\mathrm{U}(1)_X$ will not break and the hidden-Higgs corresponds to an MCP with a quartic self-interaction. We can consider the two limiting cases:\\[0.3cm]
\noindent{\bf (i)} {$|q_{\theta} eB|\ll \mu_\theta^2$}:
\\[0.3cm]
In this case the hidden $U(1)_X$ is broken with hidden-photon mass{\footnote{The hidden-Higgs receives a mass $m_\theta^2\approx2\mu_\theta^2$.}}
\begin{align}
m_{\gamma'}^2&\approx q^2_Xg^2_X v^2_{\theta}\,.
\end{align}
As discussed earlier (cf.~Fig.~\ref{interaction}), the hidden-Higgs and any other hidden matter with $\mathrm{U}(1)_X$ charge do not couple to the photon at large distances ({\it i.e.}~small momentum transfer). Hence, LSW experiments with a sufficiently low or vanishing magnetic field are sensitive to photon--hidden-photon oscillation depending on the mass differences of the propagation states~\cite{Okun:1982xi,Ahlers:2007rd,Ahlers:2007qf}. The conversion probability is given as
\begin{equation}
P_{\gamma\to\gamma'} = 4\chi^2\sin^2\left(\frac{m_{\gamma'}^2}{4\omega}z\right)\,.
\end{equation}
The total light-shining-through-the-wall probability is then,
\begin{equation}\label{transmagnetic}
P_\text{trans} = \left[
\frac{N_\text{pass}+1}{2}\right]P_{\gamma\to\gamma'}(\ell_1)P_{\gamma'\to\gamma}(\ell_2)\,,
\end{equation}
where $\ell_1$ ($\ell_2$) denotes the length of the magnetic field in front (behind) the wall.
\\[0.3cm]
\noindent{\bf (ii)} {$|q_\theta eB|\gg \mu_\theta^2$}:
\\[0.3cm]
The hidden-Higgs $\theta$ does \emph{not} acquire a vacuum expectation value and\footnote{It is difficult to talk about a `mass' for the {\emph{would-be}} Higgs in this case because the strong magnetic field explicitly breaks Lorentz invariance. One could define an effective mass for the propagation along the magnetic field direction ${m^{2}_{\parallel, h}}=-\mu^{2}_{\theta}+(2n+1)\left|q_{\theta} e B\right|\sim (2n+1)\left|q_{\theta} e B\right|$ (the `dimensional reduction' of this definition is then, in part, reflected by the existence of a whole tower of states labelled by $n$).}
\begin{equation}
m_{\gamma'}^2=0\,.
\end{equation}
Accordingly, the hidden-Higgs behaves like an MCP in this case and bounds on the hidden sector mixing parameter can be derived from the photon--hidden-photon oscillations induced by MCP loops as discussed in Ref.~\cite{Ahlers:2007rd}. In this limit the refractive index for parallel and perpendicular polarization with respect to the magnetic field reads~\cite{deltanref}\footnote{To be precise, to obtain the following expressions we have analytically continued the expressions for the absorption and refraction index of a scalar to negative mass-squareds. In the limit $|q_\theta eB|\gg \mu_\theta^2$, where there are no tachyonic modes, this is unproblematic.}
\begin{equation}
\label{refractive}
\Delta n^\theta_{\parallel,\perp} = \frac{1}{28\sqrt{3}\left(\Gamma(1/6)\right)^2}\left(\frac{3}{2}\right)^{2/3}\left(\frac{q_\theta eB}{\omega^2}\right)^{2/3}\left[(1)_\parallel,(3)_\perp\right]\,,
\end{equation}
and the absorption coefficient
\begin{equation}
\label{absorb}
 \kappa^\theta_{\parallel,\perp}=\frac{\omega}{14\left(\Gamma(1/6)\right)^2}\left(\frac{3}{2}\right)^{2/3}\left(\frac{q_\theta eB}{\omega^2}\right)^{2/3}\left[(1)_\parallel,(3)_\perp\right]\,.
\end{equation}
The transition probability after a distance $z$ is \cite{Ahlers:2007rd}
\begin{equation}
\label{transprob}
P^i_{\gamma\to\gamma'}(z)=P^i_{\gamma'\to\gamma}(z)=\chi^2[1+\exp(- \kappa_{i} z/\chi^2)-
2\exp(- \kappa_{i} z/(2\chi^2)){\cos(\Delta n_{i} \omega z/\chi^2)}]\,,
\end{equation}
where $i=\parallel,\perp$ denotes the polarization parallel or perpendicular to the magnetic field.

\begin{figure}[tb]
\begin{center}
\begin{picture}(400,270)(40,0)
\includegraphics[width=0.49\linewidth]{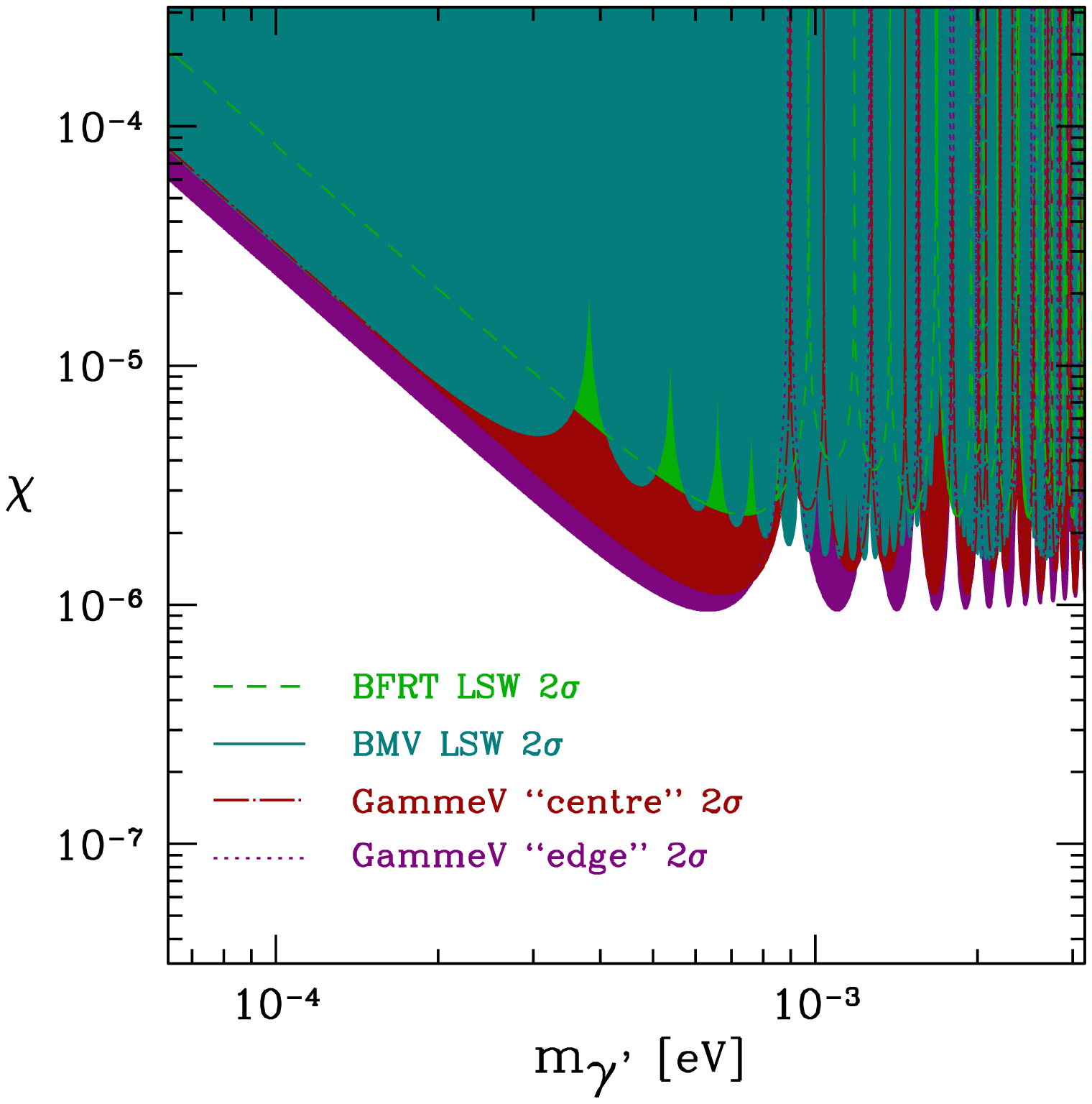}
\includegraphics[width=0.49\linewidth]{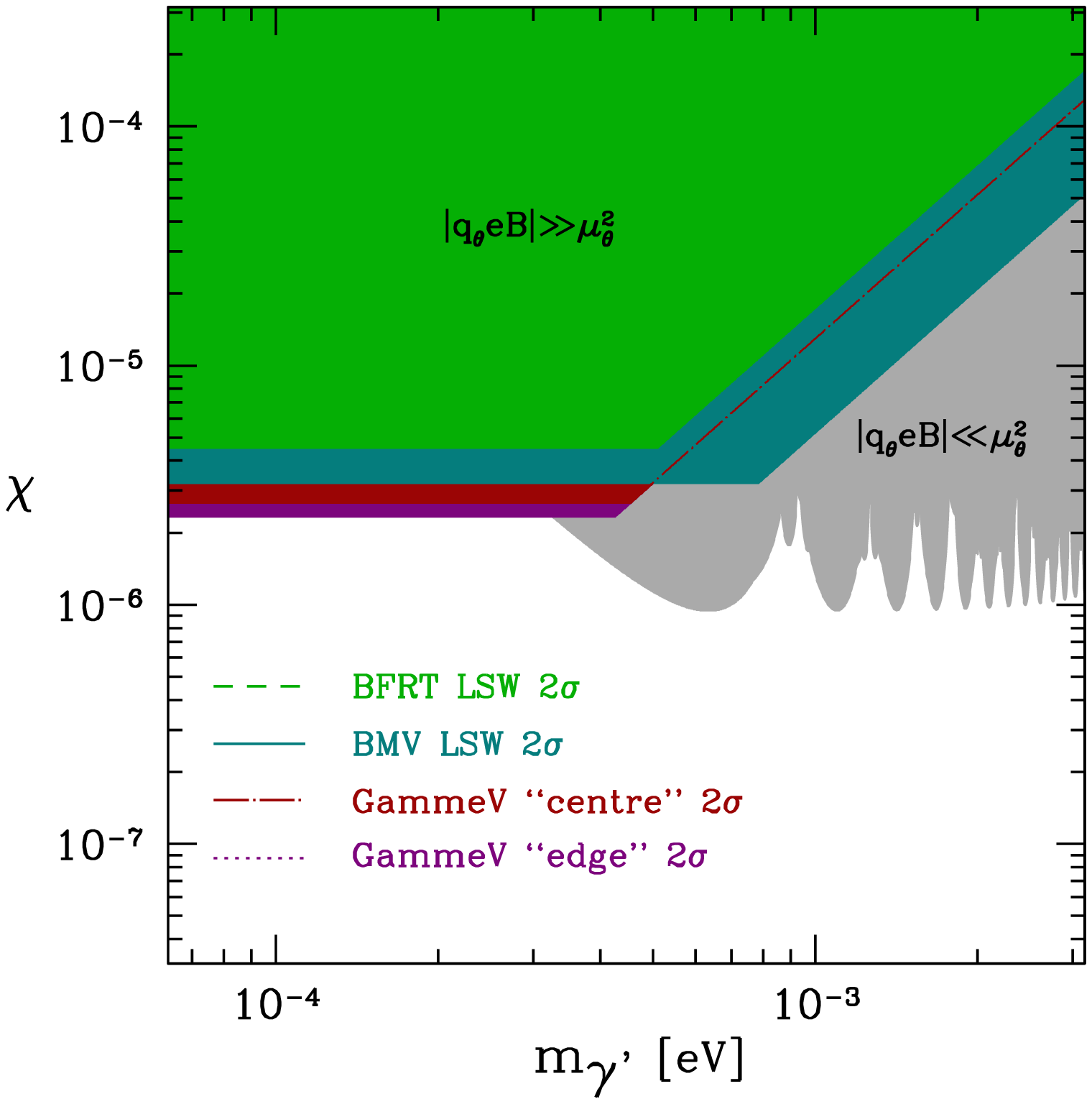}
\Text(-340,250)[c]{\scalebox{1.1}[1.1]{St\"uckelberg mechanism}}
\Text(-100,250)[c]{\scalebox{1.1}[1.1]{Higgs mechanism}}
\end{picture}
\end{center}
\vspace{-5ex}
\vspace{0.3cm}
\caption[]{\small{Upper limits on the kinetic mixing parameter as a function of the hidden-photon mass from the non-observation of light-shining-through-the-wall in the experiments BFRT, BMV and GammeV. {\bf Left Panel}: Hidden photon mass arising via the St\"uckelberg mechanism. See Ref.~\cite{Ahlers:2007qf} for details. {\bf Right Panel}: Hidden photon mass arising via the Higgs mechanism. For strong magnetic fields $|q_\theta eB|\gg \mu_\theta^2$ the hidden $\mathrm{U}(1)$ is unbroken and LSW bounds from scalar MCP loops  (cf.~\cite{Ahlers:2007rd}) apply. In this region, $m^2_{\gamma'} =  (q_Xg_X\mu_\theta)^2/\lambda_\theta$ corresponds to the mass the hidden-photon would have \emph{in vacuum}. Inside the magnetic field the hidden-photon mass is zero. For $|q_\theta eB|\ll \mu_\theta^2$ we have LSW bounds from photon-hidden-photon {oscillations arising from the mass term (grey area). We use the benchmark point $q_X= 1/2$, $g_X=e$, $\kappa=0$, and $\lambda_\theta=1$.\footnotemark[8]}. \label{sketch}}}
\end{figure}

The combined bounds {\bf (i)} and {\bf (ii)} are shown in Fig.~\ref{sketch} (right panel) and compared to the case where the hidden-photon mass is obtained via the St\"uckelberg mechanism (left panel). It is apparent that the laboratory limits on the kinetic mixing parameter can be significantly improved if the strength of the magnetic fields is sufficient to restore the hidden $\mathrm{U}(1)_X$ symmetry. Note, that in this case the
hidden-photon is \emph{massless} \emph{inside the magnetic field} and the parameter $m^{2}_{\gamma^{\prime}}=  (q_Xg_X\mu_\theta)^2/\lambda_\theta$ corresponds to the mass of the hidden-photon \emph{in vacuum}.

Finally, if one finds a positive signal in a light-shining-through-the-wall experiment one might also wonder how to distinguish between a mass arising from a Higgs- and one from a St\"uckelberg mechanism. If the mass is sufficiently small, say $m_{\gamma^{\prime}}\lesssim {\rm meV}$, one can typically achieve magnetic field strength such that $|q_\theta eB|\ll \mu_\theta^2\sim m_{\gamma^{\prime}}^2$. If the hidden-photon mass arises from a
Higgs mechanism the LSW signal will depend on the strength and the orientation (with respect to the laser polarization) of the magnetic field as can be seen from Eqs.~\eqref{refractive}, \eqref{absorb} and \eqref{transprob}. Indeed for very small $m_{\gamma^{\prime}}$ one would expect that the signal (nearly) vanishes upon switching off the magnetic field. This would not be the case if the hidden-photon mass arises from a St\"uckelberg mechanism. Moreover, for a hidden-Higgs mechanism one would expect effects in optical experiments that measure changes in the polarization when light passes through a (strong enough) magnetic field. Again no such effects are expected from a St\"uckelberg mass term.

\addtocounter{footnote}{1}\footnotetext{Strictly speaking, since the GammeV measurements were only performed with a strong magnetic field this bound is only applicable for $|q_\theta eB|\ll \mu_\theta^2$. Accordingly one should cut away a part of the first bump from the left of the grey area where this condition is  not satisfied.}

\begin{table}[tb]
\begin{center}\renewcommand{\arraystretch}{2.3}\small
\begin{tabular}{|cc|cc|}
\hline
{\normalsize Vertex}&{\normalsize Coupling}&{\normalsize Vertex}&{\normalsize Coupling}\\
\hline
$H$-$Z$-$Z$&$\displaystyle 2\frac{m_Z^2}{v_\phi}g^{\mu\nu}c_\alpha$&$h$-$Z$-$Z$&$\displaystyle -2\frac{m_Z^2}{v_\phi}g^{\mu\nu}s_\alpha$\\
$H$-$W^+$-$W^-$&$\displaystyle 2\frac{m_W^2}{v_\phi}g^{\mu\nu}c_\alpha$&$h$-$W^+$-$W^-$&$\displaystyle -2\frac{m_W^2}{v_\phi}g^{\mu\nu}s_\alpha$\\
$H$-$\gamma'$-$\gamma'$&$\displaystyle 2\frac{m_{\gamma'}^2}{v_\theta}g^{\mu\nu}s_\alpha$&$h$-$\gamma'$-$\gamma'$&$\displaystyle 2\frac{m_{\gamma'}^2}{v_\theta}g^{\mu\nu}c_\alpha$\\
$H$-$\gamma'$-$Z$&$\displaystyle 2\chi s_W \left(\frac{s_\alpha}{v_\theta}-\frac{c_\alpha}{v_\phi}\right)m_{\gamma'}^2g^{\mu\nu}$&$h$-$\gamma'$-$Z$&$\displaystyle 2\chi s_W \left(\frac{c_\alpha}{v_\theta}+\frac{s_\alpha}{v_\phi}\right)m_{\gamma'}^2g^{\mu\nu}$\\
$\bar{f}$-$\gamma^{\prime}$-$f$ & $-\chi Q_{\rm EM} e\gamma^\mu$&$\bar{f}$-$h$-$f$&$\displaystyle \frac{m_f}{v_\phi}s_\alpha$\\[0.3cm]
\hline
\end{tabular}
\end{center}
\caption[]{\small Leading order couplings for $\chi\ll1$ and $\rho\ll1$ of SM Higgs (H) and hidden-Higgs (h) to $\gamma'$, $Z$, and $W$. The last row shows also the {coupling of the Standard Model fermions $f$ to the
(mostly) hidden-photon like mass eigenstate $\gamma^{\prime}$ and the hidden-Higgs}.}\label{tab1}
\end{table}

\section{Higgs Portal Mixing}\label{Portal}

We will consider now the case that the hidden-Higgs $\theta$ mixes with the the Standard Model Higgs $\phi$ via a term $\kappa|\phi|^2|\theta|^2$~\cite{Foot:1991bp}, sometimes called the \emph{Higgs Portal}~\cite{portalref}. The particular nomenclature originates from the fact that the renormalizable coupling $\kappa$ may be large, {\it i.e.}~unsuppressed by any mass scale and enhances interactions between the Standard Model and the hidden sector. The phenomenological importance of this term has also been realized in so-called \emph{Hidden Valley} models~\cite{valleyref}.

The most general gauge-invariant and renormalizable Higgs potential (see {\it e.g.}~\cite{valleyref,higgspotref,Chang:2006fp,Chang:2007ki,Gopalakrishna:2008dv} for earlier studies) is of the form
\begin{equation}
V_\text{Higgs} = -\mu^2_\phi|\phi|^2 -\mu_\theta^2|\theta|^2+\lambda_\phi|\phi|^4+\lambda_\theta|\theta|^4+\kappa|\phi|^2|\theta|^2\,,
\end{equation}
with $\lambda_{\phi/\theta}>0$ and $4\lambda_\phi\lambda_\theta>\kappa^2$. The role of this potential in the context of gauge kinetic mixing scenarios has also been previously studied in Refs.~\cite{valleyref,Chang:2006fp,Chang:2007ki,Gopalakrishna:2008dv}. The mass of the hidden-photon and the hidden-Higgs are calculated in Appendix \ref{HiggsPortalappendix}.

In the case of a small kinetic mixing and a small ratio $\rho = v_\theta/v_\phi$ of the vacuum expectation values the hidden-Higgs mass can be expressed as
\begin{equation}\label{mh}
{m_h^2\approx m_H^2\sin^2\alpha\left(\frac{4\lambda_\theta\lambda_\phi}{\kappa^2}-1\right)}\,.
\end{equation}
After symmetry breaking, the remaining real $\theta$ and $\phi$ Higgs states mix via the Higgs Portal term. The transformation to mass eigenstates consisting of a light Higgs $h$ and the heavy Standard Model Higgs $H$ can be expressed by a rotation with an angle
\begin{equation}\label{sinalpha}
\sin\alpha \approx \frac{\kappa \rho}{2\lambda_\phi}\,.
\end{equation}
We can relate the hidden-photon mass to the hidden-Higgs mass via
\begin{equation}\label{mg}
m^{2}_{\gamma'} \approx
g^{2}_{X}q^{2}_{X}v_{\theta}\approx\frac{g_X^2q_X^2}{2\lambda_\theta}\left(m_h^2+m_H^2\sin^2\alpha\right)\,.
\end{equation}

The mass and coupling of the hidden-photon is not affected by the portal term to leading order (as long as we keep $v_{\theta}$ fixed). The Higgs self-interaction and the coupling to matter is analogous to the Standard Model with the substitution\footnote{Here, we haven't introduced hidden matter with couplings to the hidden-Higgs, that would also contribute to the
interactions of Higgs and hidden-Higgs after rotation.} $H\to(c_\alpha H-s_\alpha h)$. In Tab.~\ref{tab1} we show the leading order couplings of Higgs and hidden-Higgs to $\gamma'$, $Z$, and $W$ for a small kinetic mixing $\chi$ and ratio $\rho$ of the vacuum expectation values. Note, that in the absence of Portal mixing, $\kappa=0$, the inter-sector coupling between the Higgs and gauge bosons shown in the first two rows of Tab.~\ref{tab1} and $f$-$h$-$f$ vanish, but due to the kinetic mixing the $h$-$\gamma'$-$Z$, $H$-$\gamma'$-$Z$, and $\bar f$-$\gamma'$-$f$ cross terms shown in the third and forth row will still be present.

In particular, the $h$-$\gamma'$-$Z$ coupling could contribute to the decay with of $Z$ as has been pointed out in~\cite{Chang:2007ki}. However, we would like to emphasize that typical detectors of high-energy collider experiments contain a strong magnetic field. In the case of $|\chi q_Xg_XB|\gg\mu^2_\theta-2\kappa v^{2}_{\phi}$ (cf.~Sect.~\ref{magnetic}) the hidden-Higgs will not develop a vacuum expectation value and the hidden-photon mass together with the  $h$-$\gamma'$-$Z$ coupling vanishes. The contribution of the hidden-Higgs to the muon's anomalous magnetic moment has also been discussed in Ref.~\cite{Chang:2007ki}. For a hidden-Higgs much lighter than the muon the contribution to $(g-2)$ is given by $\Delta a_\mu^{h} \approx 10^{-8}\sin^2\alpha$~\cite{Gunion:1989we}. With $\Delta a_\mu^\text{exp} - \Delta a_\mu^\text{SM} = \text{few}\times10^{-9}$~\cite{Yao:2006px} this gives only a very mild bound on $\sin\alpha$.

We will discuss in the following two sections further possibilities how the Higgs Portal mixing with a light hidden-Higgs may be constrained by laser polarization and fifth force experiments.

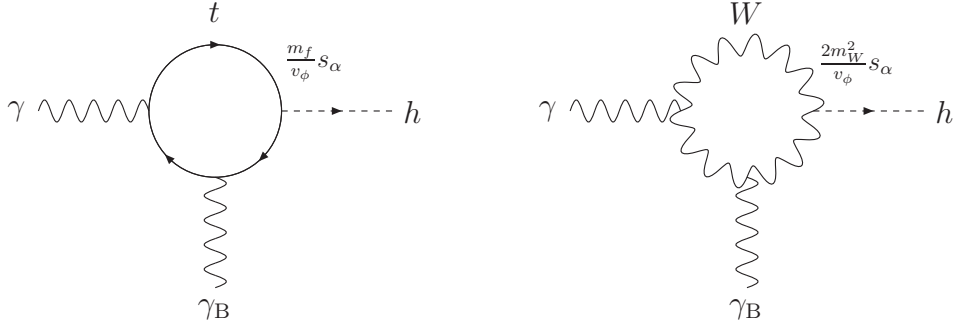
\begin{figure}[tb]
\begin{center}
\hspace*{0.8cm}
\scalebox{0.833}[0.833]{
\begin{picture}(190,140)(30,0)
\Photon(30,90)(80,90){5}{4.5}
\ArrowArcn(110,90)(30,0,90)
\ArrowArcn(110,90)(30,90,180)
\ArrowArcn(110,90)(30,180,360)
\Photon(110,60)(110,10){5}{4.5}
\Text(20,90)[c]{\scalebox{1.2}[1.2]{$\gamma$}}
\Text(110,0)[c]{\scalebox{1.2}[1.2]{$\gamma_{{\rm B}}$}}
\Text(110,135)[c]{\scalebox{1.2}[1.2]{$t$}}
\Text(155,112)[c]{\scalebox{1.}[1.]{$\frac{m_f}{{v_\phi}}s_\alpha$}}
\DashArrowLine(140,90)(190,90){3}
\Text(200,90)[c]{\scalebox{1.2}[1.2]{$h$}}
\end{picture}
}
\hspace{1cm}
\scalebox{0.833}[0.833]{
\begin{picture}(190,140)(30,0)
\Photon(30,90)(80,90){5}{4.5}
\PhotonArc(110,90)(30,0,360){5}{14}
\Photon(110,60)(110,10){5}{4.5}
\Text(20,90)[c]{\scalebox{1.2}[1.2]{$\gamma$}}
\Text(110,0)[c]{\scalebox{1.2}[1.2]{$\gamma_{{\rm B}}$}}
\Text(110,135)[c]{\scalebox{1.2}[1.2]{$W$}}
\Text(160,112)[c]{\scalebox{1.}[1.]{$\frac{2m_W^2}{{v_\phi}}s_\alpha$}}
\DashArrowLine(140,90)(190,90){3}
\Text(200,90)[c]{\scalebox{1.2}[1.2]{$h$}}
\end{picture}
}
\end{center}
\caption[]{\small {Diagrams contributing to photon--hidden-Higgs oscillations} in an external magenetic field.}\label{fig1}
\end{figure}

\subsection{Higgs Portal in Laser Polarization Experiments}

\begin{figure}[tb]
\begin{center}
\includegraphics[width=0.48\linewidth,clip=true]{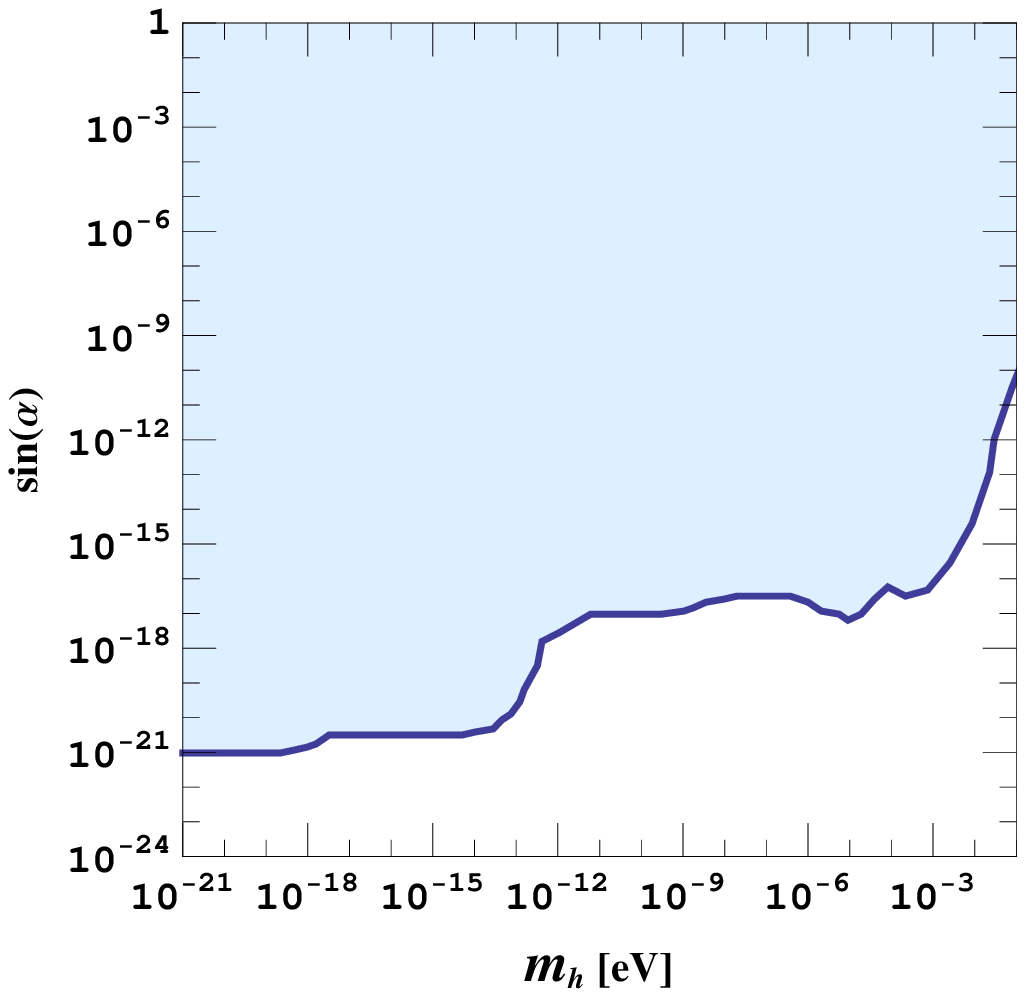}
\hfill
\includegraphics[width=0.48\linewidth,clip=true]{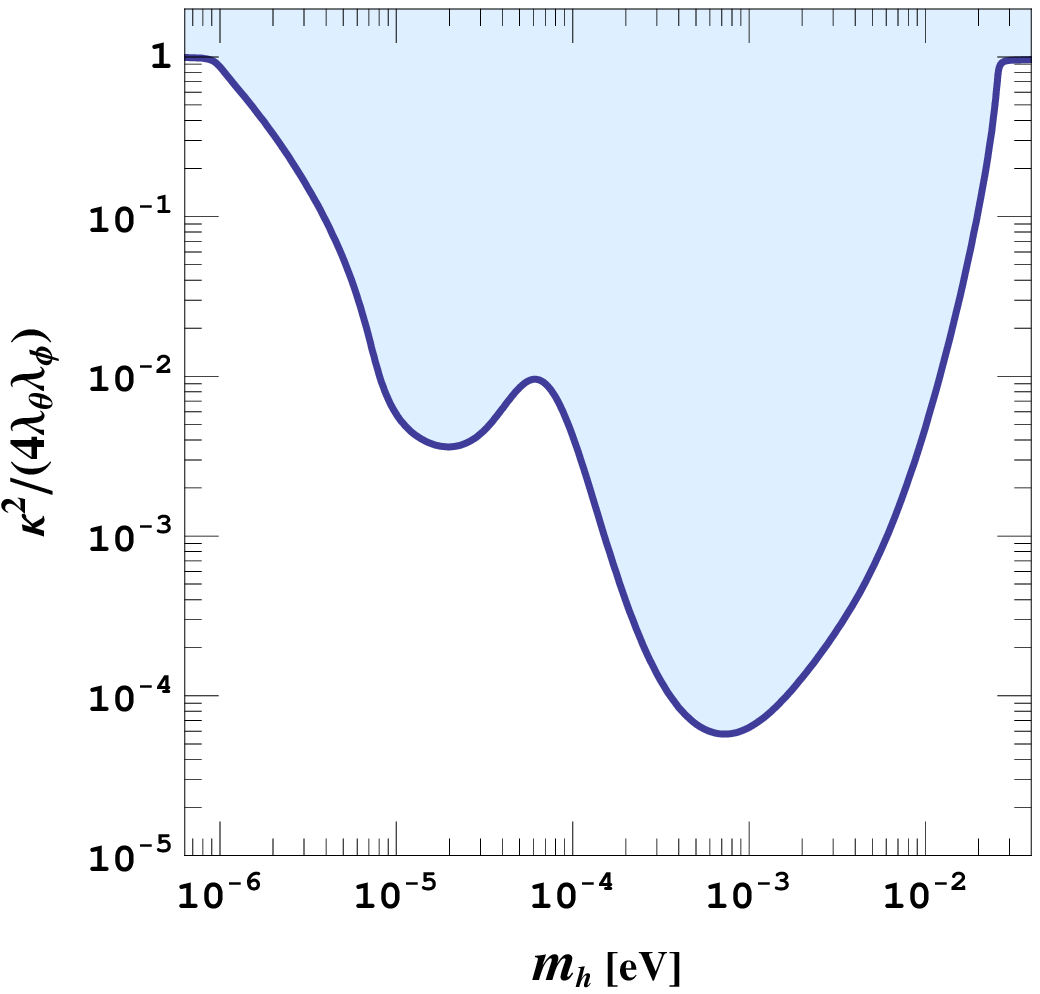}
\end{center}
\vspace{-0.3cm} \caption[]{\small Bounds on the Higgs--hidden-Higgs mixing $\sin\alpha$ from fifth force experiments as a function of the hidden-Higgs mass $m_{h}$ (left panel). Using a Higgs mass of $m_{H}=120\,{\rm GeV}$ we can translate these constraints into bounds on the scaled Higgs Portal term $\kappa^2/(4\lambda_\phi\lambda_\theta)$ as a function of the hidden-Higgs mass (right panel). For the latter we get interesting constraints in the $\mu{\rm eV}$ to $0.1$~eV range. Note, that the hidden-Higgs mass is related to the hidden-photon mass by Eq.~(\ref{mg}) and symmetry breaking of the hidden $\text{U}(1)_X$ requires $\kappa^2/(4\lambda_\phi\lambda_\theta)<1$.}\label{nonnewton}
\end{figure}

The physical hidden-Higgs in the broken phase does not directly couple to the photon. However, there can be strong loop effects~\cite{Gunion:1989we} in presence of a Higgs Portal coupling. The contribution to the hidden-Higgs coupling to two photons via a top quark and $W$ loops is shown in Fig.~\ref{fig1}. The effective Lagrangian of the hidden-Higgs photon interactions is given by an axion-like term
\begin{equation}\label{effL}
\mathcal{L}_{h\gamma\gamma} = \frac{1}{4}g_{h\gamma\gamma}hF^{\mu\nu}F_{\mu\nu}\,,
\end{equation}
where $F^{\mu\nu}$ is the field tensor of the photon and
\begin{equation}\label{effcoupling}
g_{h\gamma\gamma} = \frac{7\alpha}{3\pi v_\phi}s_\alpha \simeq2.2\times10^{-5}\text{GeV}^{-1}s_\alpha\,.
\end{equation}
Recent laser polarization and regeneration experiments are sensitive to light axion-like particles with $m_h\lesssim1\,\text{eV}$ and couplings $g_{h\gamma \gamma}\gtrsim{\rm few}\times 10^{-7}\ {\rm GeV}^{-1}$. Accordingly the null result observed in these experiments~\cite{Zavattini:2007ee,Chou:2007zz,Afanasev:2008jt} gives a bound of the order of $s_{\alpha}\sim 10^{-2}$. In the next subsection we will, however, find much stronger bounds: for sub-eV hidden-Higgs masses mixing angles $\gtrsim 10^{-10}$ are excluded by fifth force tests (cf. Fig.~\ref{nonnewton}).

\subsection{Non-Newtonian Forces}

If a very light hidden-Higgs exists, it may signal its existence through a spin-independent non-Newtonian
contribution to the gravitational force~\cite{Dupays:2006dp}. In order to derive the
corresponding bound on the hidden-Higgs parameters, we have to calculate the hidden-Higgs--nucleon--nucleon couplig $g_{\theta N N}$ appearing in the low energy effective Lagrangian
\begin{equation}
{\mathcal L} = - g_{\theta N N}\, \theta\, \overline{\psi}_N \psi_N\,.
\end{equation}
The corresponding calculation for the Standard Model Higgs $\phi$ has been done in Ref.~\cite{Shifman:1978zn} and nicely reviewed in chapter 2 of Ref.~\cite{Gunion:1989we}. Exploiting low-energy theorems, one finds
\begin{equation}
g_{\phi N N}  = \frac{m_N}{v_\phi}\,\frac{2n_H}{3\left(11-\frac{2}{3}n_L\right)} \,,
\end{equation}
with $n_H=4$ the number of heavy quarks and $n_L=2$ the number of light quarks. The contribution of the Higgs Portal term gives
\begin{equation}
g_{\theta N N}^{(\kappa )}\approx g_{\phi N N}s_\alpha\,.
\end{equation}
Bounds on the mixing $\sin\alpha$ and on the Higgs-Portal term $\kappa$ from fifth force measurements~\cite{kapner} are shown in Fig.~\ref{nonnewton}.

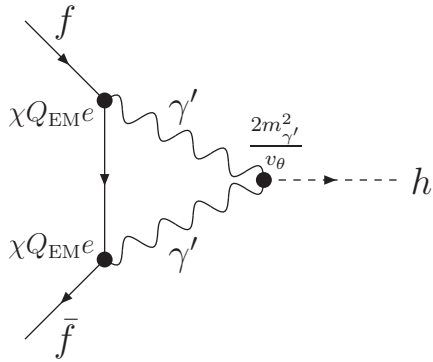
\begin{figure}[t]
\begin{center}
\scalebox{1.0}[1.0]{
\begin{picture}(250,150)(-60,-20)
\ArrowLine(0,120)(30,90) \ArrowLine(30,90)(30,30)
\ArrowLine(30,30)(0,00) \Photon(30,90)(90,60){4}{4.5}
\Photon(30,30)(90,60){-4}{4.5} \DashArrowLine(90,60)(140,60){3}
\Text(60,88)[c]{\scalebox{1.2}[1.2]{$\gamma^{\prime}$}}
\Text(60,32)[c]{\scalebox{1.2}[1.2]{$\gamma^{\prime}$}}
\Text(15,120)[c]{\scalebox{1.2}[1.2]{$f$}}
\Text(15,0)[c]{\scalebox{1.2}[1.2]{$\bar{f}$}}
\Text(150,60)[c]{\scalebox{1.2}[1.2]{$h$}}
\Text(10,85)[c]{\scalebox{0.9}[0.9]{$\chi Q_{{\rm EM}} e$}}
\Text(10,35)[c]{\scalebox{0.9}[0.9]{$\chi Q_{{\rm EM}} e$}}
\Vertex(30,90){3} \Vertex(30,30){3} \Vertex(90,60){3}
\Text(95,75)[c]{\scalebox{1}[1]{$\frac{2m^{2}_{\gamma^{\prime}}}{v_{\theta}}$}}
\end{picture}
}
\end{center}
\vspace{-1.0cm} \caption[]{\small {Kinetic mixing contribution to the coupling of a Standard Model fermion $f$, {\it e.g.}~quarks or electrons to the hidden-Higgs (the couplings for the \emph{mass eigenstates} can be found in Tab.~\ref{tab1}). The quark coupling in turn contributes to the coupling to the nucleon.}}
\label{nucleonmixing}
\end{figure}

In addition, we have also couplings of the hidden-Higgs to nucleons but also to the electrons arising from the kinetic mixing.
The leading order contribution\footnote{The coupling of a massive hidden-photon to fermions with strength $g_{\gamma^\prime ff}^{(\chi )} =\chi Q_{\rm EM}e$ does not significantly contribute to non-Newtonian forces, since the coupling is mass independent and contributions from protons and electrons largely cancel. Instead, deviations from Coulomb's law can be tested.} from the kinetic mixing correspond to triangle diagrams as shown in Fig.~\ref{nucleonmixing} with hidden-photons in the loop. The hidden-photon coupling (after rotation) to a fermion is $\chi Q_{\rm EM} e$ and the coupling to the hidden-Higgs is $2 m_{\gamma^\prime}^2/v_\theta$ (see Tab.~\ref{tab1}). For small mixings we have {$m^{2}_{\gamma^\prime} = q_\theta^2 g_X^2 v_\theta^2$} and the coupling can be estimated as
\begin{equation}
g_{\theta ff}^{(\chi )} = ({\rm loop\ factor})\,\times\, (\chi Q_{\rm EM} e)^2\,\times\, \left(\frac{2 q_{X} g_X m_{\gamma^{\prime}}}{m_f}\right)\,.
\end{equation}
Since the coupling is inversely proportional to the mass the largest contribution arises from electrons. Using a loop factor of the order of $\sim 1/(16\pi^2)\sim 10^{-2}$ we find
\begin{equation}
g_{\theta ee}\lesssim 10^{-22}\quad{\rm for}\,\, m_{\gamma^{\prime}}\lesssim {\rm meV} \,\,{\rm and}\,\, \chi\lesssim 10^{-5}\,.
\end{equation}
Fifth force measurements are sensitive to $g\sim  {\rm few}\times 10^{-22}$. Therefore, at the moment, this coupling does not yield a stronger bound. One can check that this is also true for masses $m_{\gamma^{\prime}}\gtrsim {\rm meV}$.

%

\section{Conclusions}\label{conclusions}

\begin{figure}[tb]
\begin{center}
\includegraphics[width=0.7\linewidth]{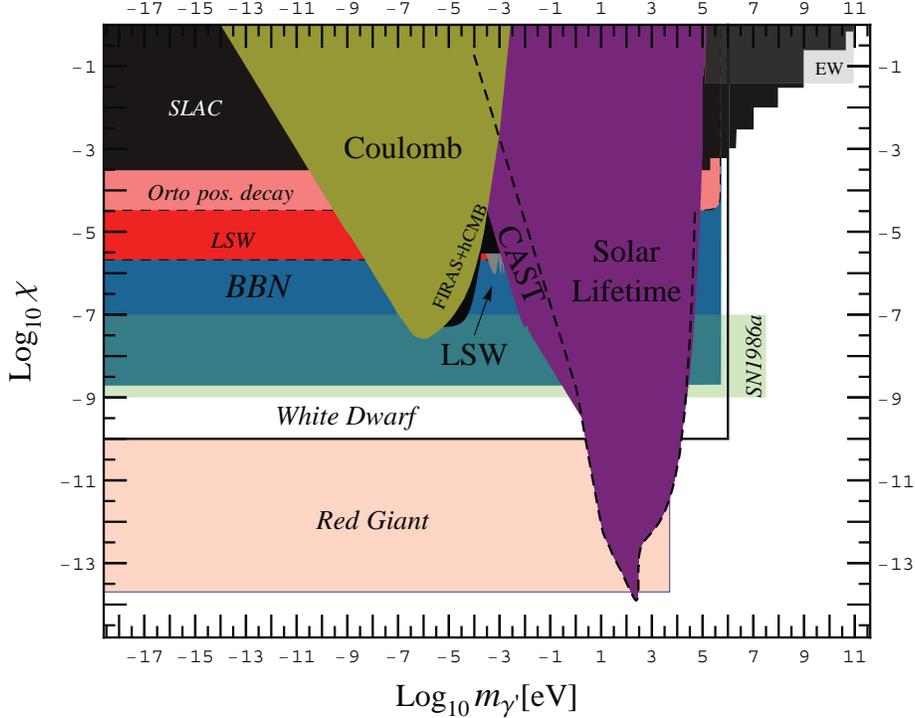}
\end{center}
\vspace{-2ex} \caption[...]{\small Bounds on the kinetic mixing parameter for massive hidden sector photons (cf. the caption of Fig.~\ref{bounds1}).
Regions labeled in italic are the the bounds that apply if the mass arises from a Higgs mechanism, and the Higgs boson appears as a minicharged particle. We have bounds from a SLAC beamdump experiment, invisible orthopositronium decays, light-shining-through-walls experiments (LSW), big bang nucleosynthesis (BBN), and energy loss considerations in supernovae~(SN1986a), white dwarfs and red giants (see~\cite{Davidson:1993sj,Badertscher:2006fm,Davidson:2000hf,Ahlers:2007qf}). Notice that we have assumed $g_Xq_X=e$ (so $|\chi|=|q_\theta|$) and $m_{\gamma'}\simeq m_h$. For $g_X q_X \ll e$, typically $m_h\gg m_{\gamma'}$ and the new bounds move upwards and to the left.\label{newbounds}}
\end{figure}

Extra `hidden' U(1) gauge bosons appear in many extensions of the Standard Model. The bounds on these hidden sector photons depend crucially on their mass. This mass can arise either via a St\"uckelberg mechanism or from a Higgs mechanism. In this paper we have investigated if one can use knowledge about the mechanism that generates the mass to improve the bounds. In particular, we have focused on the case of the Higgs mechanism. The crucial point in the case of the Higgs mechanism is that it provides an extra degree of freedom which leads to additional experimental and observational constraints. Indeed, at large momentum transfer as, {\it e.g.}, in the interior of stars, a light hidden-Higgs behaves as a minicharged particle. A similar behavior is found inside strong magnetic fields. This can be used to translate bounds on minicharged particles into bounds on massive hidden sector photons. As can be seen from Fig.~\ref{newbounds} this leads to a dramatic strengthening of the (astrophysical as well as laboratory) bounds for small masses.

The hidden-Higgs field $\theta$ also provides new potential couplings to the Standard Model. In particular, it allows for a renormalizable interaction with the Standard Model Higgs $\phi$ via a so-called Higgs-Portal term $\kappa|\theta|^2|\phi|^2$. This coupling leads to fifth-force type couplings which can be used to obtain bounds on this coupling which are independent of the size of the kinetic mixing between photon and hidden-photon (cf.~Fig.~\ref{nonnewton}). This opens the `Higgs-Portal' to the physics of light hidden sectors.

\section*{Acknowledgements}

M.~A.~acknowledges support by STFC UK (PP/D00036X/1).


\begin{appendix}

\section{Hidden-Higgs and Hidden-Photon Masses}\label{HiggsPortalappendix}

We assume that the two Higgs fields, $\phi$ and $\theta$,
are representations $(2,1/2,0)$ and $(1,0,q_{X})$ under $\mathrm{SU}(2)\times\mathrm{U}(1)_Y\times\mathrm{U}(1)_X$, respectively. The most general gauge-invariant and renormalizable Higgs potential for the Standard Model Higgs $\phi$ and the hidden-Higgs $\theta$ is of the form
\begin{equation}
V_\text{Higgs} = -\mu^2_\phi|\phi|^2 -\mu_\theta^2|\theta|^2+\lambda_\phi|\phi|^4+\lambda_\theta|\theta|^4+\kappa|\phi|^2|\theta|^2\,,
\end{equation}
with $\lambda_{\phi/\theta}>0$ and $4\lambda_\phi\lambda_\theta>\kappa^2$. The spontaneous breaking of both $\mathrm{U}(1)$'s requires
\begin{align}
\langle\phi\rangle^2&=\frac{v_\phi^2}{2} = \frac{2\lambda_\theta \mu_\phi^2-\kappa \mu_\theta^2}{4\lambda_\phi\lambda_\theta-\kappa^2}>0\,,
&\langle\theta\rangle^2&=\frac{v_\theta^2}{2} = \frac{2\lambda_\phi \mu_\theta^2-\kappa \mu_\phi^2}{4\lambda_\phi\lambda_\theta-\kappa^2}>0\,.
\end{align}
These conditions are satisfied if $\mu_\phi^2/\kappa > \mu_\theta^2/2\lambda_\theta$ and $\mu_\theta^2/\kappa > \mu_\phi^2/2\lambda_\phi$.

\subsection{Hidden-Higgs Mass}

After symmetry breaking the mass terms of the two real Higgs, emerging via the replacements\footnote{{We are using unitary gauge.}}
\begin{align}
\theta &\to \frac{1}{\sqrt{2}}\left(v_\theta+\theta\right)\,,&\phi&\to \frac{1}{\sqrt{2}}\left(\begin{matrix}0\\v_\phi+\phi\end{matrix}\right)\,,
\end{align}
is given by
\begin{equation}
\mathcal{L}_\text{Higgs\,mass} = -\frac{1}{2}\left(\,\theta\,\,\phi\,\right)\begin{pmatrix}2\lambda_\theta v_\theta^2&\kappa v_\phi v_\theta\\\kappa v_\phi v_\theta&2\lambda_\phi v_\phi^2\end{pmatrix}\begin{pmatrix}\,\theta\,\\\,\phi\,\end{pmatrix}\,.
\end{equation}
This can be diagonalized by a rotation\,,
\begin{equation}
\begin{pmatrix}h\\H\end{pmatrix} = \begin{pmatrix}\cos\alpha&-\sin\alpha\\\sin\alpha&\cos\alpha\end{pmatrix}\begin{pmatrix}\,\theta\,\\\,\phi\,\end{pmatrix}\,,
\end{equation}
with mixing angle $\tan(2\alpha) = \kappa\rho/(\lambda_\phi -\lambda_\theta \rho^2)$ and ratio $\rho = v_\theta/v_\phi$ of the two vacuum expectation values. The hidden-Higgs and Higgs mass eigenstates are given as
\begin{equation}
m^2_{h/H} = \lambda_\phi v_\phi^2+\lambda_\theta v_\theta^2\mp\sqrt{\kappa^2 v_\phi^2 v_\theta^2+(\lambda_\phi v_\phi^2-\lambda_\theta v_\theta^2)^2}\,.
\end{equation}
If we consider a small ratio of the Higgs VEVs $\rho$, we arrive at the leading order expressions
\begin{align}\label{eq1}
m_H^2&\approx 2v_\phi^2\lambda_\phi\,,&m_h^2&{\approx m_H^2\sin^2\alpha\left(\frac{4\lambda_\theta\lambda_\phi}{\kappa^2}-1\right)}\,,&\sin\alpha&\approx\frac{\kappa \rho}{2\lambda_\phi}\,.
\end{align}

\subsection{Hidden-Photon Mass}

After spontaneous symmetry breaking the gauge bosons receive a mass term
\begin{equation}
\mathcal{L}_\text{U(1)\,mass} = \frac{1}{2}\frac{v_\phi^2}{4}\left(A^3_\mu\,B_\mu\,X_\mu\right)\begin{pmatrix}g^2&-gg'&0\\-gg'&g'^2+\chi^2\widetilde g_X^2\rho^2&-\chi\widetilde g_X^2\rho^2\\0&-\chi\widetilde g_X^2\rho^2&\widetilde g_X^2\rho^2\end{pmatrix}\begin{pmatrix}A^{3\mu}\\B^\mu\\X^\mu\end{pmatrix} \,,
\end{equation}
with $\widetilde g_X = 2q_\theta g_X$. Besides the massless solution corresponding to the photon the massive modes are given by
\begin{align}
m^2_{Z}+m^2_{\gamma'} &\equiv M^2 = \frac{v_\phi^2}{4}\left(g^2+g'^2+\widetilde g_X^2\rho^2(1+\chi^2)\right)\,,\\
m^2_{Z}-m^2_{\gamma'} &\equiv \Delta^2=\sqrt{M^4-\frac{v_\phi^4}{4}\widetilde g_X^2\rho^2(g'^2+g^2(1+\chi^2))}\,.
\end{align}
These expressions reduce to the well-known expression of the Z-boson mass in the absence of $\mathrm{U}(1)_X$ breaking $\rho\to 0$. The W-boson mass is simply $m_W^2 = g^2v_\phi^2/4$.
The electromagnetic coupling after symmetry breaking is given as
\begin{equation}
e = \frac{gg'}{\sqrt{g'^2+g^2(1+\chi^2)}} = \frac{gs_W}{\sqrt{1+\chi^2c_W}}\,,
\end{equation}
with charge $Q_Y+T_3$ and the usual definition of the mixing angle. Note, that all particles that are initially only charged under $U(1)_X$ do not couple to the photon after symmetry breaking due to their alignment with the hidden-Higgs field.

In the following we are going to assume that either the kinetic mixing effect is small, {\it i.e.}~$\chi\ll1$, or that the $\mathrm{U}(1)_X$ coupling $g_X$ is small compared to $g$. In this case the electroweak mixing between $A^3$ and $B$ is given by the usual expressions $c_W\approx g/\sqrt{g'^2+g^2}$ and \mbox{$s_W\approx g'/\sqrt{g'^2+g^2}$}. We also assume that the hidden-photon mass is smaller than the $Z$ boson mass (wee {\it e.g.}~Ref.~\cite{Chang:2007ki}), which to leading order in $\chi$ or $g_X$ is the usual expression
\begin{equation}
m_Z^2 \approx \frac{g^2+g'^2}{4}v_\phi^2\,.
\end{equation}
Then to leading order in $\chi$ and $g_X$ the hidden-photon mass is given by
\begin{equation}
m_{\gamma'}^2\approx q_X^2g_X^2 v_\theta^2\,.
\end{equation}

\end{appendix}


\end{document}